\documentclass[10pt,A4paper]{article}

\usepackage{changepage}

\usepackage[utf8]{inputenc}

\usepackage{textcomp,marvosym}

\usepackage{fixltx2e}

\usepackage{amsmath,amssymb}

\usepackage{nameref,hyperref}

\usepackage{rotating}
\usepackage{natbib}

\raggedright
\usepackage[aboveskip=1pt,labelfont=bf,labelsep=period,justification=raggedright,singlelinecheck=off]{caption}

\makeatletter
\renewcommand{\@biblabel}[1]{\quad#1.}
\makeatother

\date{}

\usepackage{lastpage,fancyhdr,graphicx}
\usepackage{epstopdf}

\newcommand{\beginsupplementtext}{%
		\setcounter{table}{0}
        \renewcommand{\thetable}{S\arabic{table}}%
        \setcounter{figure}{0}
        \renewcommand{\thefigure}{S\arabic{figure}}%
        \setcounter{equation}{0}
        \renewcommand{\theequation}{S\arabic{equation}}%
     }

\begin{document}
\vspace*{0.35in}

\begin{flushleft}
{\Large
\textbf\newline{Mechanosensing in myosin filament solves a 60 years old conflict in skeletal muscle modeling between  high power output and  slow rise in tension}
}
\newline
\\
Lorenzo Marcucci\textsuperscript{1,2*\Yinyang\ddag},
Carlo Reggiani\textsuperscript{1,\ddag},
\\
\bigskip
\bf{1} Department of Biomedical Sciences, Padova University, Padova, Italy
\\
\bf{2}  Quantitative Biology Center, RIKEN, Suita, Japan
\\
\bigskip

%
%
\Yinyang This author contributed to the conception and design of the model, the numerical simulations, and analysis of results

\ddag These authors contributed to interpretation of results, drafting and critical revision of the article.




* lorenzo.marcucci@gmail.com

\end{flushleft}
\section*{Abstract}
Almost 60 years ago Andrew Huxley with his seminal paper \cite{Huxley1957} laid the foundation of modern muscle modeling, linking chemical events to mechanical performance. He described mechanics and energetics of muscle contraction through the cyclical attachment and detachment of myosin motors to the actin filament with ad hoc assumptions on the dependence of the rate constants on the strain of the myosin motors. That relatively simple hypothesis is still present in recent models, even though with several modifications to adapt the model to the different experimental constraints which became subsequently available. However, already in that paper, one controversial aspect of the model became clear. Relatively high attachment and detachment rates of myosin to the actin filament were needed to simulate the high power output at intermediate velocity of contraction. However, these rates were incompatible with the relatively slow rise in tension after activation, despite the rise should be generated by the same rate functions. This discrepancy has not been fully solved till today, despite several hypotheses have been forwarded to reconcile the two aspects.
Here, using a conventional muscle model, we show that the recently revealed mechanosensing mechanism of recruitment of myosin motors \cite{Linarietal2015} can solve this long standing problem without any further ad-hoc hypotheses.

\section*{Author Summary}
Muscle contraction is generated by the cyclical interaction of myosin motors with the actin filament.  In muscle models, relatively high attachment and detachment rates of myosin motors are needed to simulate the high power output at intermediate velocity of contraction. However, these rates are incompatible with the relatively slow rise in tension after activation. This conflict, surviving since the first model based on molecular mechanism proposed by A.F. Huxley on 1957, has been previously solved only assuming ad hoc hypotheses on the actin or myosin properties. Here we show that incorporating a recently revealed mechanosensing mechanism of recruitment of myosin motors in a simple conventional muscle model can solve this long standing problem without any further hypotheses.


\section*{Introduction}
Muscle contraction is generated by the cyclical interaction of myosin motors with the actin filament. Following attachment the myosin motor produces force and relative filaments sliding. This cyclical interaction has been extensively studied and characterized leading to the definition of the myosin-actin cycle. The kinetics of the transitions between the myosin states in the cycle are responsible for both the isometric contraction, when, following activation, the force rises to a maximum steady value, and the isotonic contraction, when the muscle shortens at a constant velocity which increases with the reduction of the load, according to the force-velocity relation \cite{Hill1938}. Muscle models based on relatively simple myosin-actin cycles, show that the kinetics needed to simulate the high-power output at intermediate shortening velocities, imply a rate of force development in isometric contraction much faster than that experimentally observed \cite{PL1995,HT1997,Mansson2010,CMD2013,MY2012}.
Despite decades of experimental discoveries and characterization of a more rich environment in the cross bridge cycle, apparently no simple explanation is able to reconcile this controversial behavior. Different additional mechanisms have been introduced on the myosin-actin cycle to solve the problem: either a non-conventional path in cross-bridge cycle \cite{PL1995}, or an active role of the second myosin head \cite{HT1997}, or a dependence of the rate of attachment on the shortening velocity \cite{Mansson2010}, or the possibility that one \cite{CMD2013} or more actin monomers become available for the same attached myosin during one ATPase cycle \cite{MY2012}. None of these hypotheses, however, provided a definite convincing solution of the problem. \\
A recent breakthrough on thick filament structure has substantiated the existence of two distinct configurations of the myosin motors in the detached state. In one configuration, motors at rest lie along the thick filament backbone, regularly packed in quasi-helical grooves, as originally demonstrated with cryo-electromicroscopy \cite{Padronetal2005}, while in the second configuration they diffuse towards the thin filament. The presence in relaxed skinned fibers of two populations of myosin motors, one with a low ATP hydrolysis rate (super-relaxed or OFF state) and the other, more disordered (active or ON state) with ten times higher ATPase rate, has been first shown by Cooke in relaxed skinned fibers \cite{SFCC2010}. The muscle activation leads to a transition of the myosin motors from the OFF state to the ON state. Only in this latter state, myosins can bind the actin filaments and perform the power stroke that leads to generation of force and/or filament displacement. Recently published evidence on intact frog muscle fibers at low temperature, shows that the rise of filament stress accompanying high load contraction induces progressive recruitment of motors for the interaction with the actin \cite{Linarietal2015}. At zero or very low force (less than one tenth of the maximum isometric force $T_0$), only a very few myosin motors are in the ON state, i.e. ready to bind actin filament, and are responsible for the rapid development of the ability to shorten at the maximum velocity \cite{LM1984}. The number of motors in the ON state increases during the rise of isometric force and reaches the maximum value at about 0.5 $T_0$.
In support to the stress dependent transition, keeping to zero the tension of a contracting muscle fiber with the imposition of shortening at the maximum velocity $V_0$ for $20-40$ milliseconds, sensibly reduces the population in the active state, favoring the transition from the ON to the OFF state. The change in the proportion between the two populations, affects the time required for the recovery of a steady tension when the isometric condition is restored. The time constant for the rise in tension after initial activation, thus starting from almost all myosins in the OFF state, has been estimated at about $\tau_R=34$ ms. After an unloaded period of 40 ms, corresponding to a shortening of 10\% of the muscle length, the tension recovery is faster, with a time constant of $\tau_{10}=28$ ms, because only a part of the ON myosins turns OFF. After an unloaded period of 20 ms, corresponding to a shortening of 5\% of the muscle length, the time constant is even lower, $\tau_5=24$ ms, because an even minor part of myosins became OFF \cite{Linarietal2015}. In our view, this imply that the characteristic time scale of the ON-OFF transition is, then, in the same order of magnitude of the attachment detachment process. Consequently, this force regulator may be responsible for the relatively slower rise after activation compared to the relatively faster attachment and detachment rates required for the high power output at intermediate velocities of contraction. \\
In the present work we test the above hypothesis introducing the experimentally observed mechanosensing mechanism into a recently designed, conventional model of one of the authors \cite{MWY2016}. At the best of our knowledge, this is the first time that such a mechanism is embedded in  the cross-bridge cycle governing a muscle contraction model. 
The previous model is able to reproduce a number of experimental data, but it shares the common limit of mathematical models described before: the maximum power output is relatively small compared to the experimental data, because of a relatively high inflection in the force velocity curve (Fig. 4 in \cite{MWY2016}). 
Starting from that conventional model, we imposed the observed thick filament tension dependence on the rate constants between the two non- force generating states, here described in term of OFF state and ON state.
This modification introduces the mechanosensing mechanism. 
The description of all other stable states are kept the same. We then compare the simulated behavior to highlight the physiological meaning of the  observed property.
\section*{Material and Methods}
\begin{figure}
\begin{center}
\includegraphics[width=.95\textwidth,natwidth=956,natheight=685]{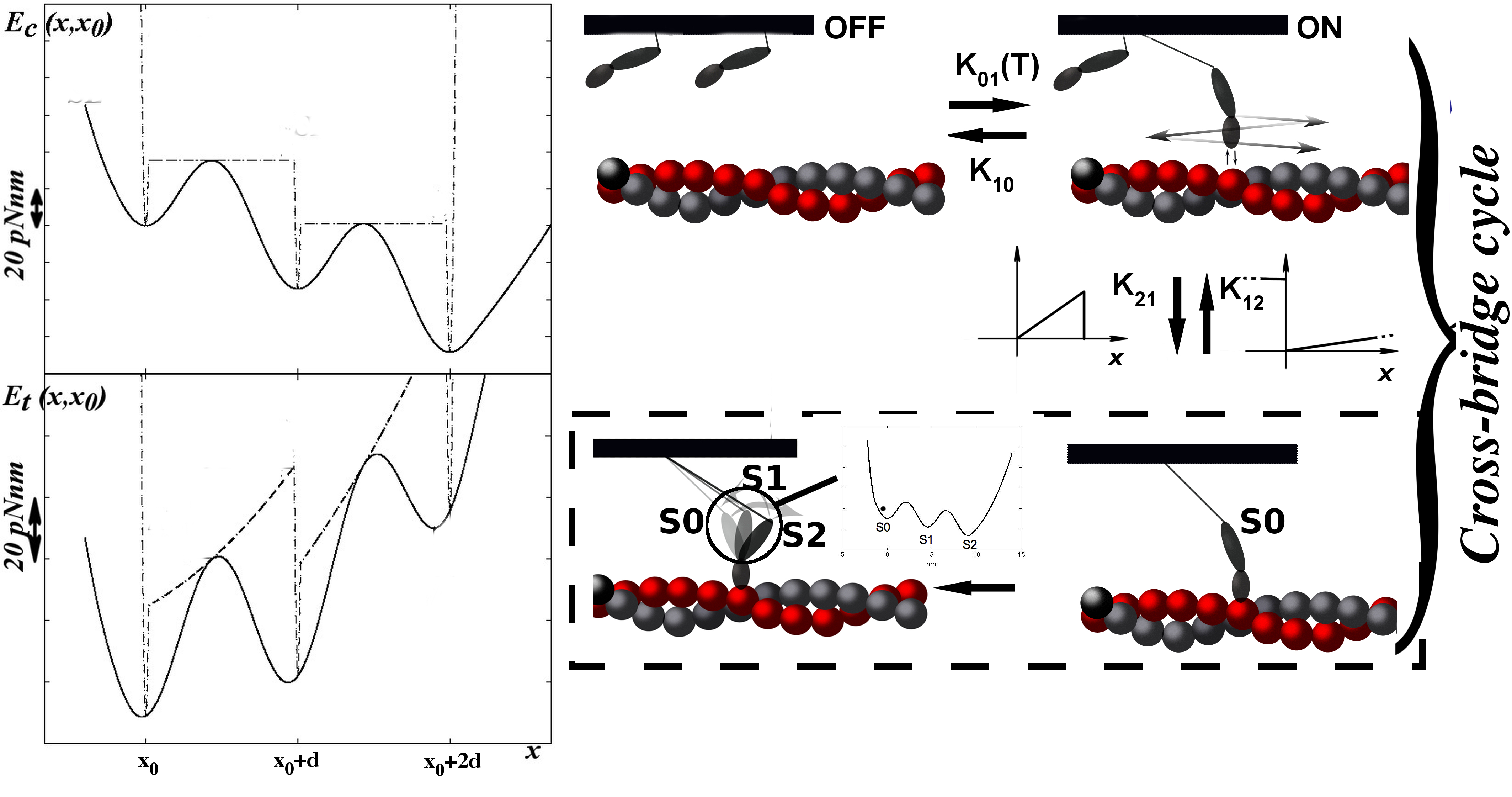}
\caption{
MS model. The model is based on a conventional model \cite{MWY2016}  with the inclusion of the mechanosensing mechanism in the rate constants between the two non-force generating states. The conventional model differs only for the absence of the mechanosensing mechanism and the inclusion of the classic actin-based cooperativity (see Matherial and Methods and SI). Rates for the attachment to and detachment from strongly attached states are functions of the strain in the $i-th$ motor ($x_i$), following the original dependence hypothesized by A.F. Huxley \cite{Huxley1957}. The power stroke has two steps as reported in the left top panel. Actomyosin energy is biased by 8kT every step. Left bottom figure represents the actomyosin energy plus the elastic component due to the myosin stiffness. $x_0$ is the stretch at the time of attachment, zero in figure.
}\label{fig_Model}
\end{center}
\end{figure}

\begin{figure}
\begin{center}
\includegraphics[width=.5\textwidth,natwidth=956,natheight=685]{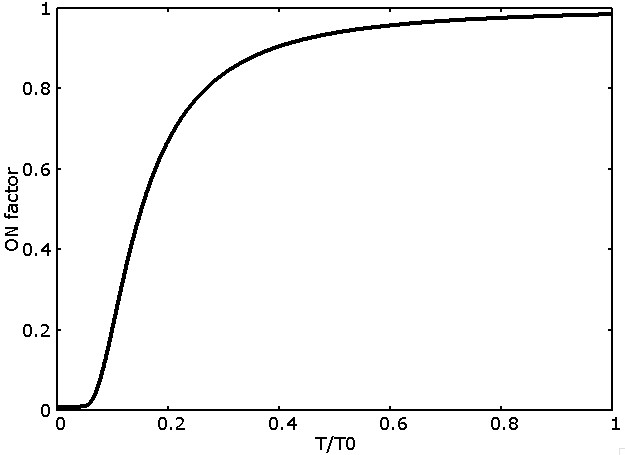}
\caption{
ON factor dependence from the relative tension on the thick filament. Experimental data refers to SM6 signal and a direct comparison with the model prediction is not possible (see text). Then, the ON factor has been defined to (i) reach about 5\% of ON motors a low forces (less than 0.1 $T_0$), and (ii) saturating at forces higher than 0.5 $T_0$.
}\label{fig_ONfact}
\end{center}
\end{figure}

In this paper we test the above mentioned hypothesis comparing the two mathematical models using a Monte-Carlo numerical method. 
We will refer to the old model as conventional model and to the new  one as mechanosensing (MS) model.  
The conventional model is described in \cite{MWY2016}, we resume its main features here and describe the modifications to include the mechanosensing mechanism. The model (see Figure \ref{fig_Model}) describes a single half-sarcomere with $N_{fil}$ couples of thin and thick filaments and $N_{xb}$ myosin motors per thick filament (compatible with the physiological values in the 2-D simplification, see Table S1 and SI). We simulate the tension-time and force-velocity curves of the whole fiber, under the hypothesis of homogeneous behavior in the series of sarcomeres.
Importantly, the experimental reference is given by the frog muscle fiber contractile behavior at $4\,^{\circ}{\rm C}$ as described in \cite{PRLea2007,Linarietal2015,LPL1992}. \\
Both filaments are rigid. Actin filaments are constrained to have the same value in the $z$-direction through the Z-line   which represents the coupling term in the mathematical model. $z=0$ refers to the optimal length of the sarcomere. 
Myosin motors are attached through an elastic element to the thick filament backbone and can cyclically interact with an actin filament. Their position is defined by $x_i$ (i=1,2...$N_{xb}$), the stretch of the motor elastic element. Myosin motor has then five stable states in both models: two non-force-generating states and three force-generating states (including a two step power stroke). In MS model, the non-force-generating states correspond to the super-relaxed, of OFF state,  with a negligible ATP hydrolysis, and to the disordered relaxed, or ON, state. In the conventional model,  they are representative of the detached and weakly attached state. For simplicity we keep the OFF and ON nomenclature. Motors in the ON state are ready to attach to actin filament forming the strongly attached states. 
In the conventional model, the OFF and ON transition rates, $k_{01}$ and $k_{10}$, respectively, include the cooperativity between myosin motors. Cooperativity affects the transition rates $k_{01}$ and $k_{10}$ through a parameter $\gamma^n$, where $n$ is 0 if there are no nearest neighbors and 1 or 2 if one or two, respectively, myosin motors are already weakly or strongly attached. Then $k_{01}=k^0_{01} \gamma^n$ and $k_{10}=k^0_{10} \gamma^{-n}$ \cite{RSTTB2003}. 
Such cooperativity is related to the displacement of the tropomyosin filament generated by an attached myosin motor, thus it is related to the thin filament.
In MS model, we  abandon the cooperativity mechanism, instead we introduce the novelty of the mechanosensing mechanism, i.e. a stress dependent regulation based on the thick filament. The OFF to ON transition rate depends on the force born by the thick filament as observed experimentally \cite{Linarietal2015}.  As described in the introduction, the probability of switching between ON and OFF depends on the tension in the thick filament.
Experimentally, the mechanosensing mechanism has been studied through the X-ray diffraction method, in particular analyzing the intensity of M6 reflection at different tensions acting on the thick backbone. It is not straightforward to make a direct relationship between that observable and the myosins distribution in the model since it would require a detailed description of the three-dimensional geometry as well as some hypotheses on the backbone elasticity which may be not linear. Then, as the simplest hypothesis, the OFF to ON rate ($k_{01}$) is a constant multiplied by a factor $ON_f(T)$, which is a function of the tension borne by the thick filament. $ON_f(T)$ is defined as $ON_f(T)=m+(1-m)exp(-t_{on}/T^2)$. The shape of this function, shown in Figure \ref{fig_ONfact}, closely mimics the obtained experimental observations (values of the parameters are reported in Supplemental Information Table \ref{supp_tab_3}).


The inverse transition rate, $k_{10}$, has been chosen constant as the simplest hypothesis. The ratio of these two rates is chosen to allow for about 5\% of myosin motors in the ON state when muscle is relaxed, as hypothesized in the experimental work. \\
In the ON state, myosins  are  in equilibrium with the strongly attached state.  
Both the conventional and MS model have three strongly attached states (S), in either pre-power stroke (S0), first (S1) or second (S2) post-power stroke. In the MS model, myosin motors detach from the strongly attached state switching to the ON state, while in the conventional model the detachment lead to a OFF state. In the OFF state, myosin motors are completely prevented to strongly attach to the actin filament.\\
From here on the models are equal. The attachment occurs only in the pre-power stroke state, while the detachment can occur in both the pre and the post power stroke state. The attachment to and detachment form the S states follow the classical H57 hypothesis \cite{Huxley1957}. The attachment rate $k_{12}$ is non-zero only in the stretched configuration of the elastic element ($x>0$), increases linearly with $x$ up to $x_{lim}$. The detachment rate, $k^+_{20}$ or $k^+_{21}$ for the conventional (S to OFF) or the MS (S to ON) model respectively, also increases linearly with $x>0$, and attains a very high value when $x<0$ ($k^-_{20}$ or $k^-_{21}$). The former hypothesis, related to the Brownian search and catch mechanism, has an original observations in Myosin V \cite{FIIMY2012}, and has been extensively used in muscle modeling under different modifications. If the absolute value of $x$ exceeds $20 \ nm$, mechanical dislodging occurs.
Numerical values used for the models are resumed in Table \ref{supp_tab_3}. \\
In both models Calcium concentration is saturated and TT units reach the full activation within the first millisecond \cite{FBSSI2014} (see SI in \cite{MWY2016}). The strongly attached state is described as a continuous energy landscape with three minima, corresponding to a pre-power stroke and a power stroke with two steps. The central region of the energy landscape is defined as:
\begin{equation}
E_c(x)=H \sin{(2 \pi x/d+\alpha_d)} + F_{atp} x
\label{sinusfunc}
\end{equation}
$x$, as said, is the stretch of the elastic element in each myosin motor, respect its anchor on the thick filament. $x_0$ is the stretch at the time of attachment, which depends on the H57 hypothesis plus a random component within [-2.5 \ nm,\ 2.5\ nm] to simulate the actin diameter \cite{HT1996}. ATP energy release favors the power stroke by biasing the biochemical energy toward the post power stroke. To simulate this bias, we added a linear drop $F_{atp}$ of $8 KT$ ($K$ is the Boltzmann constant and $T$ denotes the absolute temperature) every $d$, the distance between two consecutive minima, to the flat sinusoidal part. $\alpha_{d}$ is a constant angle that adjusts for the constant ATP drop by shifting the first minimum to $x=x_0$. Meanwhile, the convex part is expressed by a polynomial, ensuring a continuous first derivative. The energy barrier in the sinusoidal function, H, is chosen to be $5.8 KT$. We can then approximate the minimum in the energy as a parabola with stiffness $48 \ pN/nm$, much higher than the $2 \ pN/nm$ stiffness of the myosin motor. \\
The lengths of the half-actin and myosin filaments are LA=1224 nm and LB=925 nm, respectively. The myosin filament includes LB=50 nm of bare zone with no myosin motors. We included 76 myosin acting on one thin filament.
\section*{Results and Discussion}

\begin{figure}[h!]
\begin{center}
\centerline{\includegraphics[width=.5\textwidth,natwidth=3632,natheight=3580]{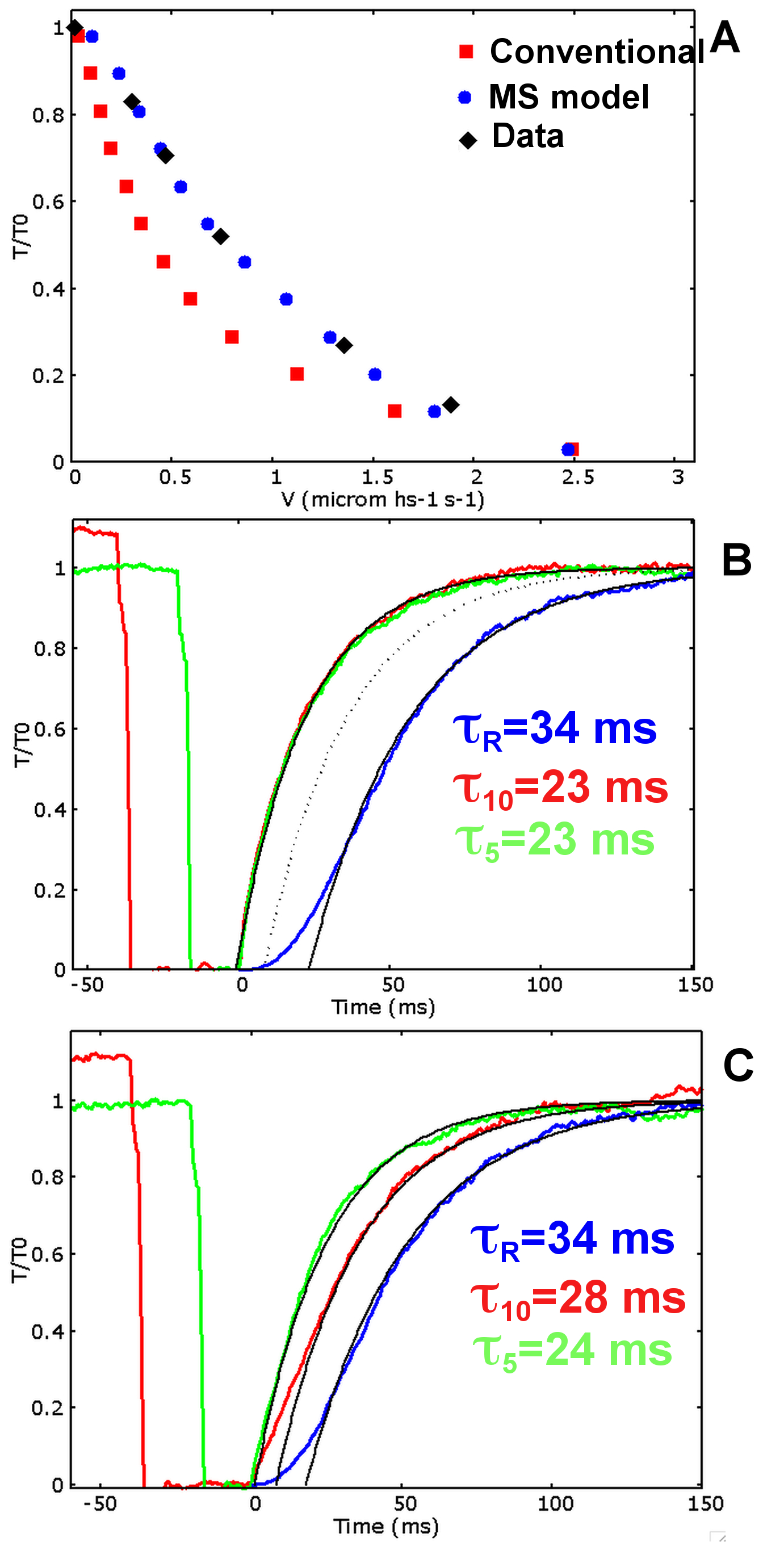}}
\caption{
Simulations for the two models. \textbf{A}: simulated and experimental data for the force velocity curves. Experimental data: filled diamond (from \cite{PRLea2007}), conventional model: red squares, MS model: blue dots. \textbf{B} Conventional model. Simulated rise in tension after initial activation (blue), and tension recovery after 40 (red) and 20 (green) milliseconds at zero tension shortening phase. The rise in tension is comparable with the experimental data, as well as the recovery after 20 ms at zero tension, but the recovery rate does not depend on the duration of the zero tension period.
\textbf{C} MS model. Simulated rise in tension after initial activation (blue), and tension recovery after 40 (red) and 20 (green) milliseconds at zero tension shortening phase. The single exponential fitting with the experimentally observed time scales well match the simulated tensions ($\tau_R=34$ ms for the rise after activation, $\tau_{10}=28$ ms after 10\% of muscle length shortening, and $\tau_{5}=24$ ms after 5\% of muscle length shortening). Also the time delay are quite similar to the experimental data \cite{Linarietal2015} (18 ms, 8 ms, 1 ms for the activation, 40 ms and 20 ms protocols respectively. Experimental data are 21 ms, 2 ms and 0 ms).
}\label{fig_F-Vel}
\end{center}
\end{figure}

\begin{figure} [h!]
\begin{center}
\centerline{\includegraphics[width=.5\textwidth,natwidth=3632,natheight=3580]{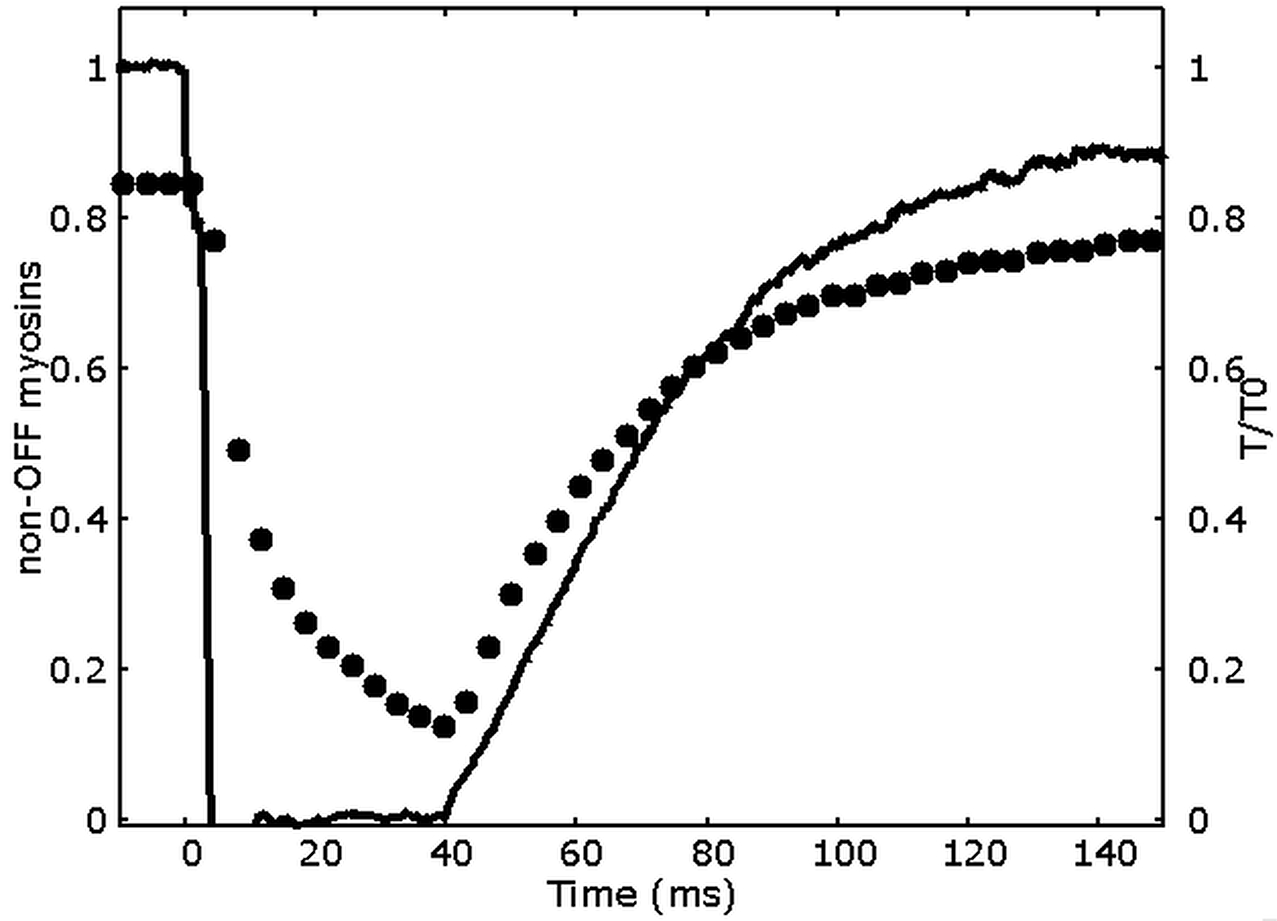}}
\caption{
Decrease in the non-OFF myosin during a 40 ms drop in tension. Force (continuos line) and relative number of non-OFF state myosins (filled circles) vs time. Non-OFF state myosins are given by the sum of ON and attached myosin motors. The behavior semi-quantitatively follow the SM6 signal.
}\label{fig_M6sim}
\end{center}
\end{figure}

\begin{figure}[h]
\begin{center}
\centerline{\includegraphics[width=.5\textwidth,natwidth=3632,natheight=3580]{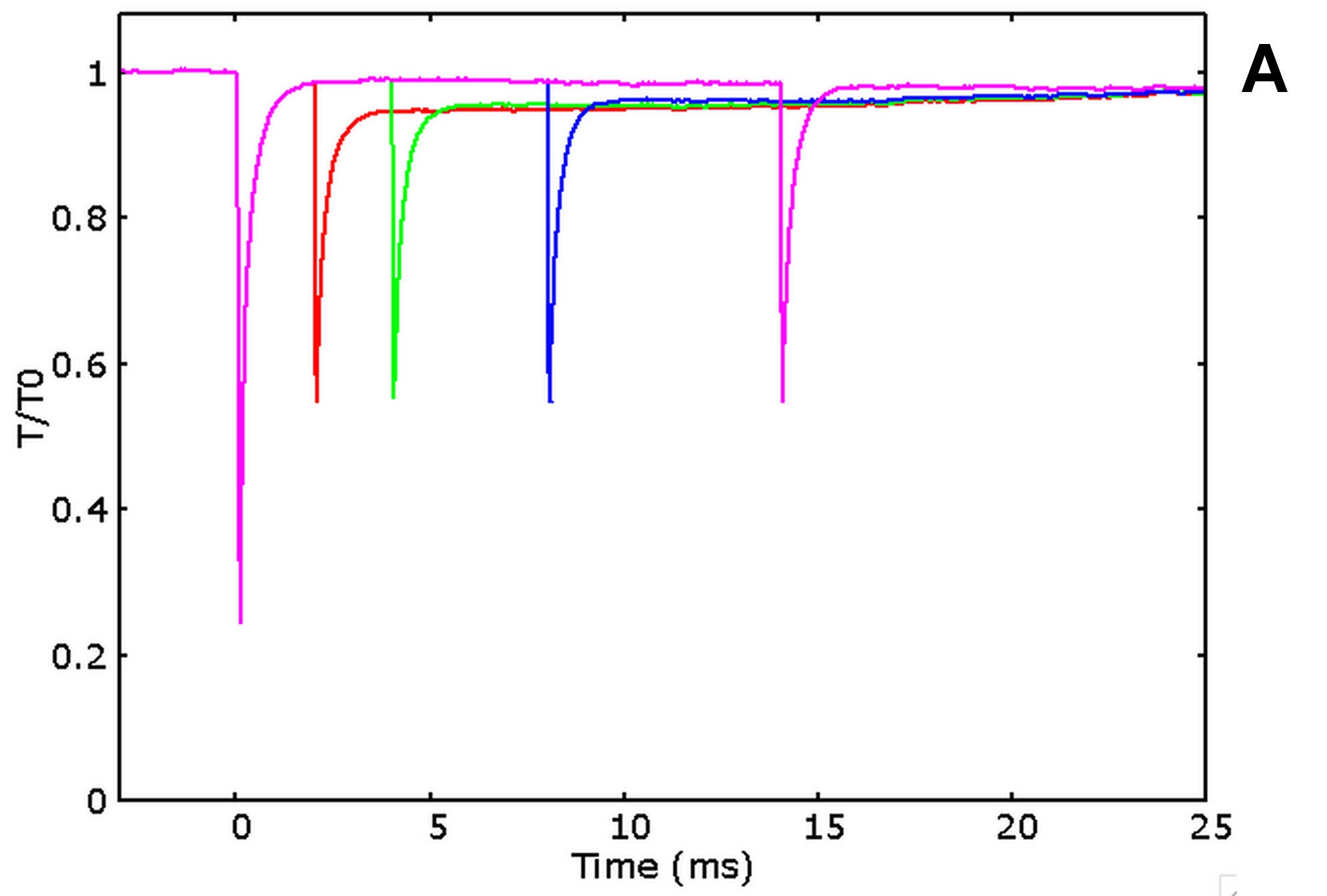}}
\caption{
Fast recovery of the power stroke. Simulated force vs. time after a conditioning length step of 5 nm and a test step of 2 nm applied after 2 ms (red), 4 ms (green), 8 ms (blue), and 14 ms (pink) after the conditioning step. The tension recovered after the test step increases with the time delayed after the conditioning step, semi-quantitatively fitting the experimental data obtained in \cite{LPL1992}.
}\label{fig_6}
\end{center}
\end{figure}

\begin{figure}[h]
\centerline{\includegraphics[width=.5\textwidth,natwidth=243,natheight=171]{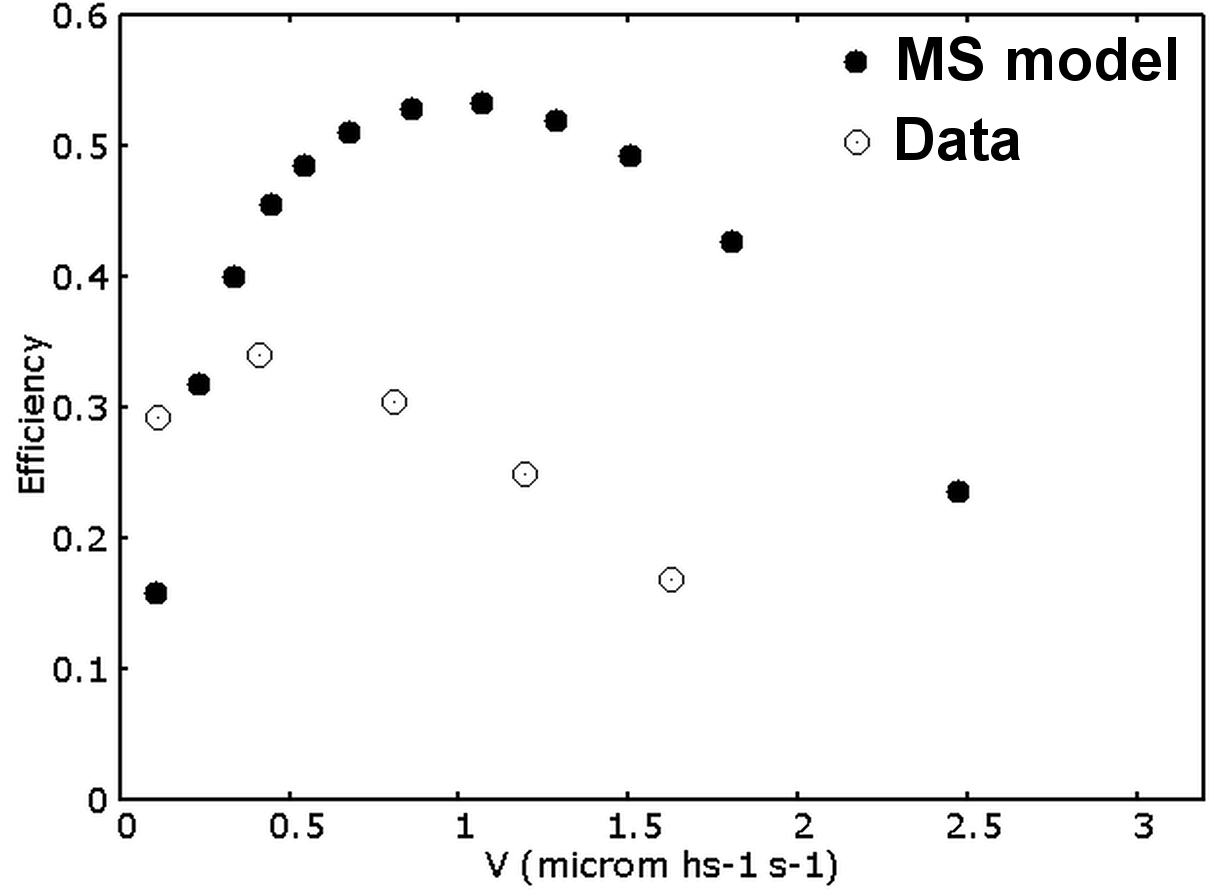}}
\caption{
Efficiency. Efficiency at different contraction velocities simulated by the MS model (filled dots) is computed as work done (force times displacement) divided by ATP consumed (85 pNnm per ATP). Values are comparable with experimental data (empty dots, from \cite{LW1995} using an half sarcomere of $1.1 \ \mu m$).
}\label{fig_5}
\end{figure}

Parameters are defined following some assumptions required to reproduce the contractile response of frog muscle fibers at low temperature. Some assumptions are shared between the two models: the time constant of the rising phase after activation must be comparable to the one observed experimentally in \cite{Linarietal2015} ($\tau_R=34$ ms); about one third of myosin motors are attached during isometric tension \cite{PRLea2007}; unloaded velocity of contraction must be comparable to experimental value.  Moreover, in the MS model we also require that in the relaxed muscle about 5 \% of myosin motors are in the ON state, as hypothesized in \cite{Linarietal2015}, and that the mechanosensing recruitment reaches the saturation at $T$ greater than $0.5 \ T_0$.
The force velocity curve simulated by the conventional model (Figure \ref{fig_F-Vel} A, red squares) shares the common limit described in the introduction: despite both the maximum velocity and the rising phase after activation (Figure \ref{fig_F-Vel}  B, blue line) are comparable with the experimental data, the velocities at intermediate forces are sensibly lower than the experimental data obtained on the same muscle type \cite{PRLea2007}, (Rana Temporaria, filled diamonds in figure). This reduces the power output in this range of forces to unrealistic low values.  \\
The limits of the conventional approach are clearly detectable when the force is allowed to recover isometrically after a period of unloaded contraction (Figure \ref{fig_F-Vel} B). The time constants, for the 20 ms (green line) and the 40 ms (red line) drop in tension, are equal to each other, as expected by a model without a feedback from the tension.  With the current parameters, the simulated time constant is comparable to the one observed for the 20 ms drop but its pretty higher than that of the 40 ms drop. A different choice of parameter would modify the outcome, but the two rates can be made different only introducing further hypotheses in the conventional model. \\
The MS model includes the stress dependent transition between the two detached states. It is able to fit excellently the force-velocity curve (Figure \ref{fig_F-Vel} A, blue dots), while keeping the proper fitting for the  rising phase during isometric activation (Figure \ref{fig_F-Vel} C, blue curve). Consequently, the power output at intermediate level of velocity of contraction is preserved at the physiological values. The model is able to keep a high maximum power thanks to a relatively high attachment and detachment rates, while the rising phase is slowed by the ON-OFF transition. These results strongly support the idea that the mechanosensing mechanism in the recruitment of ON or active myosins may solve the long standing conflict described before. \\
The simulated tension transients after unloaded contraction are shown in Figure \ref{fig_F-Vel} C. The curves for the tension development after initial activation, and after an imposed shortening at zero tension for 20 and 40 milliseconds, closely follow the single exponential fitting with time constants respectively of 34 ms, 24 ms and 28 ms. These values are the same as the ones observed experimentally \cite{Linarietal2015}, and also the time delays are quite similar (see caption of Figure \ref{fig_F-Vel}. Interestingly, the characteristic timing of the observed mechanosensing mechanism, is compatible with the attachment and detachment rate constants required to fit the high power output at intermediate velocities.
Differently to the conventional model, now we can reproduce a lower time constant after longer zero-tension periods, because an higher number of myosin motors move to the OFF state in MS model, thanks to the mechanosensing system. The ON-OFF transition can be satisfactorily simulated as model mimics the increase of the myosins in OFF state during ramp shortening, roughly following what observed experimentally (figure \ref{fig_M6sim}).  \\
It is important also to test the predictive ability of the model in relation to another behavior which has already been connected to the above discussed conflict, the so-called fast recovery of the power stroke. 
This behavior has been shown first about 25 years ago by Lombardi and co-workers \cite{LPL1992}. The tension recovery after a small "test" step in length, which follows a bigger "conditioning" step during steady state of the isometric contraction, increases with the time interval between the two steps, showing a greater recovery in tension already after 4 ms. The recovery increases more and more up to 15 ms after the first step. This evidence is incompatible with the slower time scale of the attachment detachment process, as inferred during the rising phase in conventional models. 
In Figure \ref{fig_6} are shown the simulations of the experimental protocol used for the multiple steps in \cite{LPL1992}. The model is able to almost quantitatively fit the experimental behavior, clearly showing an improved capacity to recover the original tension with the delay of few milliseconds from the conditioning step. \\
Finally, the analysis of the energetic predicted by the model requires some special considerations. As described before, in both the conventional and the MS models, the motors can detach from any attached state, also in the pre-power stroke. This event may be prevented in real muscle, or it may be possible but without the associated ATP splitting. Moreover, while the $F_{atp}$ drop imposed in $E_c$ ($16 \ KT$) is lower than total ATP energy ( about $20 \ KT$), the H57 hypothesis implies an energy consumption which is difficult to asses. Both effects affect the efficiency predicted by the model. Anyway, associating one ATP splitting for each detachment event, the simulated energetic is compatible with data in the literature (Figure \ref{fig_5}). 
\section*{Conclusions}
In this study we have shown that the tension dependent recruitment of active motors from the OFF state makes the tension development after activation less sensitive to the acto-myosin cycle rate constants and thus allows higher rate constant of attachment. This mechanism, which is substantiated by recent experimental findings \cite{Linarietal2015}, solves the long standing discrepancy between the high power output and the slow rise in tension in skeletal muscle contraction modeling. This conflict, surviving since the first model based on molecular mechanism proposed by A.F. Huxley on 1957 \cite{Huxley1957}, has been previously solved only introducing \textit{ad--hoc} hypotheses on the actin or myosin properties. We have also shown that the experimentally observed kinetics of this mechanism are quantitatively in agreement with our hypothesis. Perturbing in some way ON-OFF kinetics would allow to experimentally test this hypothesis. For instance, increasing, chemically or mechanically, the number of active motors in the relaxed muscle, would generate a faster tension rise after activation without affecting the force velocity relationship. Alternatively, perturbing the tension in the thick filament during activation would modify the number of active motors, and consequently affect the mechanical response of the muscle. \\
The ON-OFF regulation, requires a tension  feedback which is likely to be generated by a molecular component other than myosin or  actin, like titin or the C-protein. How this feedback information is given to each myosin motor has to be explored experimentally. The tension may modify the state of the non-myosin proteins in the thick filament, affecting the rates of each motor in the same way. Or the extension of the thick filament may disturb the myosin-myosin interaction in the OFF state, inducing a nearest-neighbor cooperative interaction. The latter is required to describe the steep slope in the pCa-force curve \cite{RSTTB2003} and the mehcanosensing mechanism may then contribute to the known coupling through the tropomyosin filament. Further experimental evidence are required for an explanation of the tension dependence, which is imposed on phenomenological basis in this paper. Nevertheless, in this work we have shown that including a new, experimentally well supported, structural property of the thick filament in muscle  modeling is as important as the precise characterization of actomyosin interaction. 
\section*{Acknowledgments}
We thank Prof. V. Lombardi and Prof. M. Linari for the useful discussions and for comments on a earlier version of the manuscript. L.M. work was supported by European Commission, Seventh Framework Programme (FP7/2007-2013) under Grant Agreement n°600376.

\beginsupplementtext
\part*{Supplementary Information}

\section{Supplementary text}

A Monte-Carlo approach is used in both models, where the state and position of each myosin is followed at each time step $\Delta t=10^{-6} s$. At each time step, the algorithm updates the position $x_i$ of the i-th myosin with respect to its anchor on the thick filament, considering the current position of the actin filament, the coupling factor for the myosin motors. In this way, the total tension generated is given by:
\begin{equation}
F_{Tot}= \sum_{j=0}^{N_{fil}} 
\sum_{i=0}^{N_{XB}}
k x_i
\label{eq:force}
\end{equation}
where $N_{XB}$ is the total number of myosin heads, and $N_{fil}$ is the number of filaments in the sarcomere, while $k$ is the stiffness of the myosin motor.\\
The position of the myosin head is first updated based on the position of the actin filament, $z$, in the previous time step:
\begin{equation}
x_i(t)=z(t)-z_i^a+x_i^a+n_{ps} d
\label{eq:xpos}
\end{equation}
Here $z_i^a$ and $x_i^a$ are the positions of the actin filament and the myosin pre-stretch, respectively, at the time of attachment of the $ i-th$ myosin head. $n_{ps}$ equals 0, 1 and 2 when the myosin head resides in the S0, S1 and S2 minimum, respectively. 
Having updated the position of each myosin head, the algorithm updates the position of the actin filament by the Newton–Raphson method. In each iteration, the total force is computed by Equation \ref{eq:force}.\\
In both models, rate constants in the attached state are computed through a Kramers-Smoluchovsky approximation of the corresponding Langevin equation of a diffusing, over-damped particle with drag coefficient $\eta$ in the sinusoidal energy landscape $E_c(x)+1/2k(x-x_0)^2$ (see Matherials and Methods):

\begin{equation}
\eta \dot{x}_i=-\omega_i(t) E^\prime (x_i) - k(x_i-x^i_0)+\sqrt{\eta \kappa_b T} \Gamma(t).\label{Langevin} 
\end{equation}

 $\omega$ is zero when the myosin head is detached and one when it is attached. $\Gamma(t)$ is a random term satisfying $<\Gamma(t)>=0$ and $<\Gamma(t_1),\Gamma(t_2)>=2\delta (t_1-t_2)$ (white noise).

The algorithm for the model requires a pre-step outside the time loop::
\begin{enumerate}
\item[0.]	Generate a matrix of rate constants for each stretching level between -80 nm and 20 nm through a numerical integration of the Kramers-Smoluchivsky approximation of the Langevin equation describing the motion of a particle in the defined potential energy. 
\end{enumerate}
Then:
\begin{enumerate}
\item Compute the tension acting on each thin filament and computing the new value of $k_{01}$
\item Define the new stable state of each myosin (OFF, ON, Pre or post power stroke) using the rate constants for the transitions $k_{01}$, $k_{10}$, $k_{12}$, $k_{21}$,$ k_f$ and $k_b$.
\item Update the positions of each myosin head and the actin filament, depending on the setup and external conditions, and calculate the total force. 
\item Update the position of the actin filament through an implicit method
\item Increment the time by one time step and return to Step 1 until total time is less than the prescribed
\end{enumerate}
%


\begin{table} [h]
\caption{{\bf Parameter values and their descriptions}}
\begin{tabular}
{ | l | l | l |} \hline Parameter & Value & Description \\ \hline k & 2 pN/nm & myosin stiffness \\ \hline $\eta$ & 70 pNns/nm & myosin drag coefficient\\  \hline LB & 50 nm & bare zone \\ \hline LM & 825 nm & Myosin filament length \\ 
\hline LA & $1224 \ nm$ & Actin filament length \\  \hline DT & $1 \  \mu s$ & time step \\ \hline T & $277.15 \,^{\circ}{\rm K}$ & Temperature \\ 
\hline $K$ & 0.0138 pN nm / K & Boltzmann constant  \\ \hline $k_{01}$ MS & $596.8*ON_{f} s^{-1}$ & OFF-ON rate \\ \hline 
$k_{10}$ MS & $206.5 \ s^{_1}$ & ON-OFF rate \\ \hline $k_{12}$  MS & $92.1 \ s^{-1}nm^{-1}$ & ON to attached rate \\ \hline
$k^+_{21}$  MS & $19.1 \ s^{-1} nm^{-1}$ & attached to ON rate when $x>0$ \\ \hline 
$k^-_{21}$ MS  & $490 \ s^{-1}$ & attached to ON rate when $x<0$ \\ \hline 
$k^0_{01}$ conv & $17.0 s^{-1}$ & OFF-ON rate \\ \hline
$k^0_{10}$ conv & $7500 \ s^{_1}$ & ON-OFF rate \\ \hline 
$\gamma$ conv & $40$ & cooperativity factor \\ \hline
$k_{12}$  conv & $30.0 \ s^{-1}nm^{-1}$ & ON to attached rate \\ \hline
$k^+_{20}$  conv & $6.2 \ s^{-1} nm^{-1}$ & attached to ON rate when $x>0$ \\ \hline 
$k^-_{20}$ conv  & $500 \ s^{-1}$ & attached to ON rate when $x<0$ \\ \hline 
$m$ & $0.01$  & minimum $ON_f$  \\ \hline
$t_f$ & $200 \ pN^2$ & inflection factor for $ON_f$  \\ \hline
$d$ & $4.6 \ nm$ & first and second power stroke step  \\ \hline
$F_{atp}$ & $8 \ KT/d \ pN$ & ATP bias of the energy  \\ \hline
$H$ & $5.7 \ KT \ pNnm$ & Barrier between minima  \\ \hline
\end{tabular} \label{supp_tab_3}
\end{table}
%

\clearpage
%
%
%
%
%

\begin{thebibliography}{19}
\makeatletter
\newcommand{\dinatlabel}[1]%
{\ifNAT@numbers\else\NAT@biblabelnum{#1}\hspace{2\labelsep}\fi}
\makeatother
\expandafter\ifx\csname natexlab\endcsname\relax\def\natexlab#1{#1}\fi
\expandafter\ifx\csname url\endcsname\relax\def\url#1{\texttt{#1}}\fi

\bibitem[Caremani u.\,a.(2013)Caremani, Melli, Dolfi, Lombardi und
  Linari]{CMD2013}
\dinatlabel{Caremani u.\,a. 2013} \textsc{Caremani}, M.~; \textsc{Melli}, L.~;
  \textsc{Dolfi}, M.~; \textsc{Lombardi}, V.~; \textsc{Linari}, M.:
\newblock {{T}he working stroke of the myosin {I}{I} motor in muscle is not
  tightly coupled to release of orthophosphate from its active site}.
\newblock In: \emph{J. Physiol. (Lond.)}
\newblock 591 (2013), Oct, Nr.~Pt 20, S.~5187--5205

\bibitem[Fujita u.\,a.(2012)Fujita, Iwaki, Iwane, Marcucci und
  Yanagida]{FIIMY2012}
\dinatlabel{Fujita u.\,a. 2012} \textsc{Fujita}, K.~; \textsc{Iwaki}, M.~;
  \textsc{Iwane}, A.~H.~; \textsc{Marcucci}, L.~; \textsc{Yanagida}, T.:
\newblock {{S}witching of myosin-{V} motion between the lever-arm swing and
  brownian search-and-catch}.
\newblock In: \emph{Nat Commun}
\newblock 3 (2012), S.~956

\bibitem[Fusi u.\,a.(2014)Fusi, Brunello, Sevrieva, Sun und Irving]{FBSSI2014}
\dinatlabel{Fusi u.\,a. 2014} \textsc{Fusi}, L.~; \textsc{Brunello}, E.~;
  \textsc{Sevrieva}, I.~R.~; \textsc{Sun}, Y.~B.~; \textsc{Irving}, M.:
\newblock {{S}tructural dynamics of troponin during activation of skeletal
  muscle}.
\newblock In: \emph{Proc. Natl. Acad. Sci. U.S.A.}
\newblock 111 (2014), Mar, Nr.~12, S.~4626--4631

\bibitem[Hill(1938)]{Hill1938}
\dinatlabel{Hill 1938} \textsc{Hill}, A.V.:
\newblock {{T}he heat of shortening and dynamics constants of muscles}.
\newblock In: \emph{Proc. R. Soc. Lond. B}
\newblock 126 (1938), Nr.~843, S.~136--195

\bibitem[HUXLEY(1957)]{Huxley1957}
\dinatlabel{HUXLEY 1957} \textsc{HUXLEY}, A.~F.:
\newblock {{M}uscle structure and theories of contraction}.
\newblock In: \emph{Prog Biophys Biophys Chem}
\newblock 7 (1957), S.~255--318

\bibitem[Huxley und Tideswell(1996)]{HT1996}
\dinatlabel{Huxley und Tideswell 1996} \textsc{Huxley}, A.~F.~;
  \textsc{Tideswell}, S.:
\newblock {{F}ilament compliance and tension transients in muscle}.
\newblock In: \emph{J. Muscle Res. Cell. Motil.}
\newblock 17 (1996), Aug, Nr.~4, S.~507--511

\bibitem[Huxley und Tideswell(1997)]{HT1997}
\dinatlabel{Huxley und Tideswell 1997} \textsc{Huxley}, A.~F.~;
  \textsc{Tideswell}, S.:
\newblock {{R}apid regeneration of power stroke in contracting muscle by
  attachment of second myosin head}.
\newblock In: \emph{J. Muscle Res. Cell. Motil.}
\newblock 18 (1997), Feb, Nr.~1, S.~111--114

\bibitem[Linari u.\,a.(2015)Linari, Brunello, Reconditi, Fusi, Caremani,
  Narayanan, Piazzesi, Lombardi und Irving]{Linarietal2015}
\dinatlabel{Linari u.\,a. 2015} \textsc{Linari}, M.~; \textsc{Brunello}, E.~;
  \textsc{Reconditi}, M.~; \textsc{Fusi}, L.~; \textsc{Caremani}, M.~;
  \textsc{Narayanan}, T.~; \textsc{Piazzesi}, G.~; \textsc{Lombardi}, V.~;
  \textsc{Irving}, M.:
\newblock {{F}orce generation by skeletal muscle is controlled by
  mechanosensing in myosin filaments}.
\newblock In: \emph{Nature}
\newblock 528 (2015), Dec, Nr.~7581, S.~276--279

\bibitem[Linari und Woledge(1995)]{LW1995}
\dinatlabel{Linari und Woledge 1995} \textsc{Linari}, M.~; \textsc{Woledge},
  R.~C.:
\newblock {{C}omparison of energy output during ramp and staircase shortening
  in frog muscle fibres}.
\newblock In: \emph{J. Physiol. (Lond.)}
\newblock 487 ( Pt 3) (1995), Sep, S.~699--710

\bibitem[Lombardi und Menchetti(1984)]{LM1984}
\dinatlabel{Lombardi und Menchetti 1984} \textsc{Lombardi}, V.~;
  \textsc{Menchetti}, G.:
\newblock {{T}he maximum velocity of shortening during the early phases of the
  contraction in frog single muscle fibres}.
\newblock In: \emph{J. Muscle Res. Cell. Motil.}
\newblock 5 (1984), Oct, Nr.~5, S.~503--513

\bibitem[Lombardi u.\,a.(1992)Lombardi, Piazzesi und Linari]{LPL1992}
\dinatlabel{Lombardi u.\,a. 1992} \textsc{Lombardi}, V.~; \textsc{Piazzesi},
  G.~; \textsc{Linari}, M.:
\newblock {{R}apid regeneration of the actin-myosin power stroke in contracting
  muscle}.
\newblock In: \emph{Nature}
\newblock 355 (1992), Feb, Nr.~6361, S.~638--641

\bibitem[Mansson(2010)]{Mansson2010}
\dinatlabel{Mansson 2010} \textsc{Mansson}, A.:
\newblock {{A}ctomyosin-{A}{D}{P} states, interhead cooperativity, and the
  force-velocity relation of skeletal muscle}.
\newblock In: \emph{Biophys. J.}
\newblock 98 (2010), Apr, Nr.~7, S.~1237--1246

\bibitem[Marcucci u.\,a.(2016)Marcucci, Washio und Yanagida]{MWY2016}
\dinatlabel{Marcucci u.\,a. 2016} \textsc{Marcucci}, L.~; \textsc{Washio}, T.~;
  \textsc{Yanagida}, T.:
\newblock {{I}ncluding thermal fluctuations in actomyosin stable states
  increases the predicted force per motor and macroscopic efficiency in muscle
  modelling.}
\newblock In: \emph{PLoS Comp. Biol.}
\newblock submitted (2016)

\bibitem[Marcucci und Yanagida(2012)]{MY2012}
\dinatlabel{Marcucci und Yanagida 2012} \textsc{Marcucci}, Lorenzo~;
  \textsc{Yanagida}, Toshio:
\newblock From Single Molecule Fluctuations to Muscle Contraction: A Brownian
  Model of A.F. Huxley's Hypotheses.
\newblock In: \emph{PLoS ONE}
\newblock 7 (2012), 07, Nr.~7, S.~e40042. --
\newblock URL \url{http://dx.doi.org/10.1371%2Fjournal.pone.0040042}

\bibitem[Piazzesi und Lombardi(1995)]{PL1995}
\dinatlabel{Piazzesi und Lombardi 1995} \textsc{Piazzesi}, G.~;
  \textsc{Lombardi}, V.:
\newblock {{A} cross-bridge model that is able to explain mechanical and
  energetic properties of shortening muscle}.
\newblock In: \emph{Biophys. J.}
\newblock 68 (1995), May, Nr.~5, S.~1966--1979

\bibitem[Piazzesi u.\,a.(2007)Piazzesi, Reconditi, Linari, Lucii, Bianco,
  Brunello, Decostre, Stewart, Gore, Irving, Irving und Lombardi]{PRLea2007}
\dinatlabel{Piazzesi u.\,a. 2007} \textsc{Piazzesi}, G.~; \textsc{Reconditi},
  M.~; \textsc{Linari}, M.~; \textsc{Lucii}, L.~; \textsc{Bianco}, P.~;
  \textsc{Brunello}, E.~; \textsc{Decostre}, V.~; \textsc{Stewart}, A.~;
  \textsc{Gore}, D.~B.~; \textsc{Irving}, T.~C.~; \textsc{Irving}, M.~;
  \textsc{Lombardi}, V.:
\newblock {{S}keletal muscle performance determined by modulation of number of
  myosin motors rather than motor force or stroke size}.
\newblock In: \emph{Cell}
\newblock 131 (2007), Nov, Nr.~4, S.~784--795

\bibitem[Rice u.\,a.(2003)Rice, Stolovitzky, Tu, de~Tombe und Bers]{RSTTB2003}
\dinatlabel{Rice u.\,a. 2003} \textsc{Rice}, J.~J.~; \textsc{Stolovitzky}, G.~;
  \textsc{Tu}, Y.~; \textsc{Tombe}, P.~P. de~; \textsc{Bers}, D.~M.:
\newblock {{I}sing model of cardiac thin filament activation with
  nearest-neighbor cooperative interactions}.
\newblock In: \emph{Biophys. J.}
\newblock 84 (2003), Feb, Nr.~2 Pt 1, S.~897--909

\bibitem[Stewart u.\,a.(2010)Stewart, Franks-Skiba, Chen und Cooke]{SFCC2010}
\dinatlabel{Stewart u.\,a. 2010} \textsc{Stewart}, M.~A.~;
  \textsc{Franks-Skiba}, K.~; \textsc{Chen}, S.~; \textsc{Cooke}, R.:
\newblock {{M}yosin {A}{T}{P} turnover rate is a mechanism involved in
  thermogenesis in resting skeletal muscle fibers}.
\newblock In: \emph{Proc. Natl. Acad. Sci. U.S.A.}
\newblock 107 (2010), Jan, Nr.~1, S.~430--435

\bibitem[Woodhead u.\,a.(2005)Woodhead, Zhao, Craig, Egelman, Alamo und
  Padron]{Padronetal2005}
\dinatlabel{Woodhead u.\,a. 2005} \textsc{Woodhead}, J.~L.~; \textsc{Zhao},
  F.~Q.~; \textsc{Craig}, R.~; \textsc{Egelman}, E.~H.~; \textsc{Alamo}, L.~;
  \textsc{Padron}, R.:
\newblock {{A}tomic model of a myosin filament in the relaxed state}.
\newblock In: \emph{Nature}
\newblock 436 (2005), Aug, Nr.~7054, S.~1195--1199

\end{thebibliography}


\end{document}